\documentclass[final]{elsart5}

 \usepackage{graphicx}

\usepackage{amssymb}

\begin{document}

\begin{frontmatter}

\title{Magnetic order in Graphite: Experimental
evidence, intrinsic and extrinsic difficulties\thanksref{tit1}}
\thanks[tit1]{We thank R. H\"ohne for the interest in this work and O. V. Yazyev
and M. Ziese for fruitful discussions. This work was supported by the
EU through ``Ferrocarbon", the DFG under ES 86/11 and the DAAD under
D/07/13369.}

\author[aff1]{P. Esquinazi\corauthref{cor1}}
\ead{esquin@physik.uni-leipzig.de}
\corauth[cor1]{Tel:+493419732750; fax:+493419732769}
\author[aff1]{J. Barzola-Quiquia}
\author[aff1]{D. Spemann}
\author[aff1]{M. Rothermel}
\author[aff2]{H. Ohldag}
\author[aff3]{N. Garc\'ia}
\author[aff1]{A. Setzer}
\author[aff1]{T. Butz}
\address[aff1]{Institute for Experimental Physics II, University of Leipzig, 04103 Leipzig, Germany}
\address[aff2]{Stanford Synchroton Radiation Laboratory, Stanford University, Menlo Park, CA 94025, USA}
\address[aff3]{Laboratorio de Fisica de Sistemas Peque\~nos y Nanotecnología, CSIC, 28049 Madrid, Spain}




\begin{abstract}
We discuss recently obtained data using different experimental
methods including magnetoresistance measurements that indicate the
existence of metal-free high-temperature magnetic order in graphite.
Intrinsic as well as extrinsic difficulties to trigger magnetic order
by irradiation of graphite are discussed in view of recently
published theoretical work.
\end{abstract}

\end{frontmatter}

\section{Introduction}\label{intro}

The possibility to have magnetic order at room temperature in a
system without the usual 3$d$ metallic magnetic elements has
attracted the interest of the solid state physics community in recent
years. According to recent theoretical studies
\cite{kusakabe03,duplock04,yaz07,kuma07,pis07} the graphite structure
with defects and/or hydrogen appears to be one of the most promising
candidates to find this phenomenon. Early reports in the decades of
the 80's and 90's indicate the existence of magnetic order in some
carbon-based samples (see \cite{esqui06} and Refs. therein). However,
difficulties to reproduce those results and the unclear role of
impurities casted doubts on its existence. The nowadays technique
allows us to measure the amount of magnetic impurities with high
enough accuracy especially in carbon-based materials. However,
inappropriate sample handling added to the possibility of non-simple
phase transformations of magnetic elements after sample preparation
(as in the case of magnetic fullerene \cite{makanat01,tal07}),
possible aging effects  as well as the rather small magnetic signals
of the magnetic carbon samples, make the research in this area of
magnetism rather difficult.

A few years ago  a first study of proton irradiation effects on the
magnetic properties of highly oriented pyrolytic graphite (HOPG) has
been reported \cite{pabloprl03}. After this experimental study
theoretical estimates of the magnetic moments at certain lattice
defects produced by irradiation followed
\cite{lehtinen03,lehtinen04}. The introduction of defects in the
graphite structure by irradiation should be in principle an ideal
method to test any possible magnetic order in carbon since it allows
to minimize sample handling and to estimate quantitatively the
produced defect density in the structure. The main magnetic effects
produced by proton irradiation have been reproduced in further
studies \cite{barzola1,barzola2}.  X-ray circular magnetic dichroism
(XMCD) studies on proton-irradiated spots on carbon films confirmed
that the magnetic order is correlated to the $\pi$-electrons of
carbon only, ruling out the existence of magnetic impurity
contributions \cite{ohldagl}. Apparently, the reproducibility of the
above reported effects by independent groups is not yet satisfactory.
Taking this fact into account the aim of this contribution is not
only to discuss new evidence that speaks for the existence of
metal-free magnetic order in graphite but also on the intrinsic and
extrinsic difficulties to trigger and measure its in general small
magnetic signals.

\section{Impurity measurements, Intrinsic and Extrinsic Difficulties}

\subsection{Impurities Measurements}

Taking into account old \cite{esqui06} as well as recently published
results \cite{pabloprl03,barzola1,barzola2,ohldagl,ramos07,par08} and
the amount of ferromagnetic mass one may trigger with a single-energy
proton beam (see sec.~\ref{int}) we expect ferromagnetic moments at
saturation of the order of several tens of $\mu$emu for typical
sample masses, at best. Therefore, one requires that: (a) the total
amount of magnetic impurities in the samples should be low enough to
provide a nominal ferromagnetic moment of at most one order of
magnitude smaller than the expected ferromagnetic moment at
saturation and produced by the irradiation, e.g. $\sim 5~\mu$emu. (b)
The use of a method to characterize the impurities with low enough
detection limit.

For pure Fe one needs a mass $m = 23~$ng to produce a magnetic moment
of $\simeq 5 \mu$emu. This means a volume of $\sim 3 \times
10^{-9}~$cm$^3$, i.e. a cube of $\simeq 14~\mu$m size. Taking into
account the typical graphite sample mass used for SQUID measurements,
this means to have a relative impurity concentration of $\sim
6~\mu$g/g. The characterization method used by us and called Particle
Induced X-ray Emission (PIXE) provides the needed detection limit
($\lesssim 0.1~\mu$g/g) for all typical magnetic elements embedded in
the carbon matrix.

\begin{figure}
\begin{center}
\includegraphics[width=0.5\textwidth]{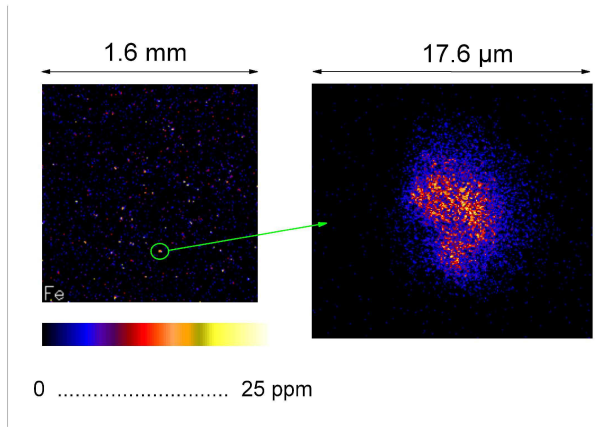}
\end{center}
\caption{(Color online) Iron distribution obtained by PIXE within
$\sim 30~\mu$m depth of a HOPG sample ($0.4^\circ$ rocking curve
width). The Fe distribution is in this case non-homogeneous, even
within a Fe spot (right picture). Rutherford backscattering (RBS)
data reveals a nearly two-dimensional Fe distribution at the spot.
The total Fe concentration in this sample is $0.6~\mu$g/g equivalent
to $\lesssim 0.15~$ppm.}\label{Fe}
\end{figure}

The main magnetic impurity in HOPG samples is Fe. Figure~\ref{Fe}
shows the Fe distribution of a typical HOPG sample. The overall Fe
content is $0.6~\mu$g/g, i.e. less than 0.15~ppm. This quantity is in
general higher than in most of the HOPG samples we have measured from
the same company (Advanced Ceramics). To check this number and the
overall calibration of our system, the impurities content of another
HOPG sample has been measured by Neutron Activation Analysis (18 days
activation time with 6~hs. measuring time) and by PIXE, both
providing the same result of $0.17 \pm 0.03~\mu$g/g of Fe. Other
magnetic impurities are clearly below this number. This means that
the whole amount of Fe, if ferromagnetic, would give a maximum
magnetic moment $< 0.5~\mu$emu. Clearly, we have to assure that the
whole sample handling does not provide additional impurities, see
Sec.~\ref{ex}.

In a recently published study the magnetic properties of HOPG samples
after implantation of Fe have been studied \cite{hoh08}. No
particular influence on the ferromagnetic properties of HOPG was
observed after up to $\sim 4000~\mu$g/g (0.08~at\%) Fe-concentration
in the implanted region. On the other hand, the implantation produced
a clear increase in the paramagnetic contribution, which is caused by
the structural disorder created by the ion bombardment.  Even a heat
treatment of the Fe-implanted HOPG samples, which reduced the
paramagnetic contribution, did not produce additional ferromagnetism
 \cite{hoh08}. The absence of a ferromagnetic
contribution after Fe-implantation indicate that iron impurities in
carbon not necessarily trigger ferromagnetism. This result also
indicates that a highly disordered, nearly amorphous carbon lattice,
as it was the case after Fe-implantation, does not show magnetic
order.

\subsection{Intrinsic Difficulties}\label{int}

(a) In spite of some differences in the  models that use limited
sample size and different algorithms for the calculations, all recent
theoretical results indicate that defects, e.g. vacancies or hydrogen
add atoms, in the graphene
\cite{kusakabe03,duplock04,yaz07,kuma07,pis07,yaz08,san08} or
graphite lattices \cite{yaz08} are a source of finite magnetic
moments and the main ingredient to trigger magnetic order. To this
consent we should add also the long-range antiferromagnetic exchange
coupling between C-neighbors belonging to the two different
sublattices of the graphene or graphite lattice
\cite{yaz07,kuma07,pis07}. An unbalance between the vacancy (or
defect) distributions in the two sublattices of graphene results in a
ferrimagnetic state  with a net magnetic moment $m \propto (N_A-
N_B)/2$, where $N_{A,B}$ is the number of vacancies in the $A$ or $B$
sublattices \cite{yaz08,san08}. The possible role of hydrogen
\cite{kusakabe03,duplock04} and the influence of the stacking order
in the graphite lattice \cite{yaz08} complicate further a realistic
estimate of the expected magnetic moment after irradiation.

Nevertheless, one expects a rather narrow window of values for
several irradiation parameters, like fluence, energy or ion current,
necessary to trigger a measurable magnetic signal. It is clear that
the selected ion fluence  will play a main role because this is one
of the parameters that determines the defect density as a function of
the penetration depth. A small enough distance between vacancies
would trigger a sufficiently large magnetic moment to be measurable
with the SQUID. On the other hand, the condition of preserving the
graphite lattice -- amorphous carbon is not ferromagnetic but
paramagnetic \cite{hohcfilms} -- provides a limit to the maximum
allowed fluence. Therefore, one expects that at a given fluence at
which, for example, the distance between vacancies is of the order of
1~nm, we will have a maximum in the magnetization of the irradiated
graphite sample, as has been observed experimentally
\cite{pabloprl03,ramos07,xia}.

\begin{figure}
\begin{center}
\includegraphics[width=0.5\textwidth]{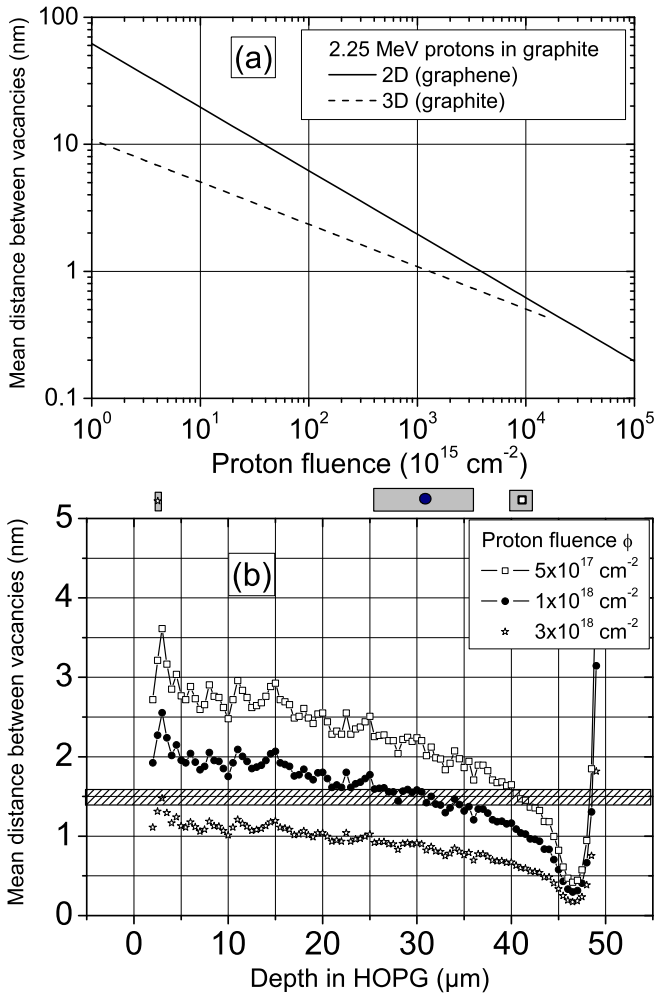}
\end{center}
\caption{(a) Mean distance between first neighboring vacancies vs.
proton fluence of 2.25 MeV energy for the two dimensional, graphene
case (straight line) and the three dimensional graphite lattice
(dashed line) assuming a vacancy production of 0.1 per micron depth
interval and incident proton \protect\cite{ziegler}. (b) Mean
distance between vacancies in the graphene case vs. depth from the
surface of a graphite sample estimated using Monte Carlo simulations
\protect\cite{ziegler} and for three fixed fluences. The shadowed
area is a guide to the eye to realize the different depths and total
sample mass ($\propto$ to the boxes drawn at the top of the figure)
at which a mean distance between vacancies of $1.5 \pm 0.1~$nm
exists.}\label{dis}
\end{figure}

The defect density depends on the penetration depth of the ions in
the sample, see Fig.~\ref{dis}. Because we can expect that the total
magnetic moment in a sample will be inversely proportional to, at
least, the square power of the distance between vacancies, at a given
penetration depth only a rather small amount of ferromagnetic mass
with large enough value of magnetization can be produced in a sample
using a fixed energy for irradiation. Below a given distance between
vacancies the crystalline order and therefore the magnetic order will
collapse. As an example how to estimate the necessary fluence range
to get the largest magnetic signal, we calculate the mean distance
between neighboring vacancies obtained from Monte Carlo simulations
of the damage produced by 2.25~MeV protons in graphite using the SRIM
program \cite{ziegler}.

Figure~\ref{dis}(a) shows the dependence of the mean distance between
vacancies with the proton fluence $\phi$ for a two-dimensional (2D)
graphene lattice, i.e. the vacancies of adjacent graphene layers are
neglected. The vacancies from adjacent graphene layers are taken into
account in the 3D graphite lattice estimate, see Fig.~\ref{dis}(a).
These estimates were obtained assuming a vacancy production rate of
0.1 vacancies per micron depth interval expected in the first $\sim
10~\mu$m depth from the graphite surface for protons of 2.25~MeV
energy. Furthermore, the depth dependency of the vacancy production
rate given by the SRIM simulation \cite{ziegler} can be used to
calculate the mean distance as a function of depth, see
Fig.~\ref{dis}(b). Assuming that the largest signal due to the
induced magnetic order is produced at a distance between vacancies of
$1.5 \pm 0.25$~nm in the 2D case, different depth intervals
contribute to the magnetic signals at different fluences. The
calculations indicate that from the three selected fluences in
Fig.~\ref{dis}(b) the largest ferromagnetic signal would be given for
$\phi \simeq 10^{18}~$protons per cm$^2$ at $\sim 30~\mu$m depth.
These estimates indicate also that the largest ferromagnetic mass one
can produce with a single energy proton beam will be located at the
first $\sim 10~\mu$m depth where the curve is rather flat.

Results obtained in Ref.~\cite{barzola2} indicate that the measured
magnetic signal coming from irradiated spots was located in the first
micrometers depth in qualitative agreement with the estimates
presented here. Following the estimates done  in Ref.~\cite{pis07}
based on the ferromagnetic contribution of coupled defects from the
same sublattice of a graphene lattice, a mean distance between
vacancies of 1.5~nm would give a critical temperature of $\sim 450~$K
in good agreement with experimental observations \cite{barzola2,xia}.
We note that due to the statistical process of vacancy production and
the simplicity of the model used for the estimate, an error of at
least a factor of two is expected for the calculated damage
production rates and the derived mean vacancy distance. Note that in
our energy range a lower proton energy will decrease the distance
between vacancies at a given depth.

(b) Other intrinsic difficulty of the irradiation method is the
temperature rise in the sample during irradiation. Ion currents of
several tens of nA can be enough to melt a glass holder as the
experience showed. A large increase of the sample temperature during
irradiation would anneal defects and increase the diffusion of
hydrogen, effects that are not taken into account in the simulations
done above. The need of low fluences increases automatically the
irradiation time (and costs) for a fixed given fluence. We note that
a clear increase of the ferromagnetic signal has been achieved after
proton irradiation at 110~K in comparison with similar irradiations
done at room temperature \cite{barzola2}.

(c) The results of Ref.~\cite{pabloprb02} showed that in general HOPG
samples before irradiation show a background ferromagnetic signal.
There is no clear experimental evidence that proves that this
background is always due to magnetic impurities. The ferromagnetic
moment (remanent or at saturation) can be of the order or even larger
than the one proton irradiation would produce, see for example
Fig.~\ref{figrema}. Therefore, the magnetic characterization of the
sample before irradiation is unavoidable to obtain the effect
produced by the ion-irradiation.

\subsection{Extrinsic Difficulties}\label{ex}

(a) Because at the moment there is no possibility to measure the
magnetic moment of a sample in the same irradiation chamber, a
minimization of sample handling is necessary. Once the sample is
measured before the first irradiation it should be transported to the
irradiation chamber and back to the SQUID holder taking care that no
additional impurities are introduced. We have designed a sample
holder that allows us to measure the magnetic moment of the sample in
the SQUID and to fix it inside the irradiation chamber without any
changes. We investigated the reproducibility of the magnetic signals
and checked that the sample handling, even after irradiating it
several times with low fluences and very small spots,
 does not produce systematic changes in the magnetic signals.
 Our arrangement provides a reproducibility of $\sim 2 \times
10^{-7}$~emu in the field range $\lesssim 2~$kOe \cite{barzola1} and
allows the subtraction of the virgin data from those after
irradiation point by point, increasing substantially the sensitivity
of the magnetic measurements to $\sim 2 \times 10^{-8}$~emu
\cite{barzola2}.

(b) SQUID measurements are time consuming. Firstly, one needs several
weeks prior to any irradiation to check for the reproducibility of
the whole procedure, including the fixing of the sample and the
transport to the irradiation chamber. The SQUID sensitivity and
reproducibility should assure that changes in the magnetic moment in
the range of $\sim 1~\mu$emu or larger are real and due to the sample
changes produced by the irradiation. The commercial SQUID apparatus
can also produce some artificial loops due in part to the hysteretic
behavior of the superconducting solenoid used to produce the magnetic
field and therefore care should be taken with very small ($\lesssim
10^{-7}~$emu) ferromagnetic-like magnetic moment signals.

(c) The last extrinsic difficulty we would like to mention is the
needed accelerator time and the estimate of the fluence. As a large
scale machine, an accelerator in the MeV range is not always
available. Therefore, one tries to minimize  the necessary amount of
irradiations. Systematic long irradiation procedures, e.g. the study
of the change of the magnetic moment vs. fluence or diameter of the
micro-spots, etc., are in general difficult to realize by a single
research group. The fluence values one provides are nominal ones
obtained after measuring the time and the (small) dc ion currents
and/or through RBS measurements and simulations. These values might
have errors due to current leaks or other electronic artifacts.

\section{SQUID measurements}\label{squid}

\begin{figure}
\begin{center}
\includegraphics[width=0.5\textwidth]{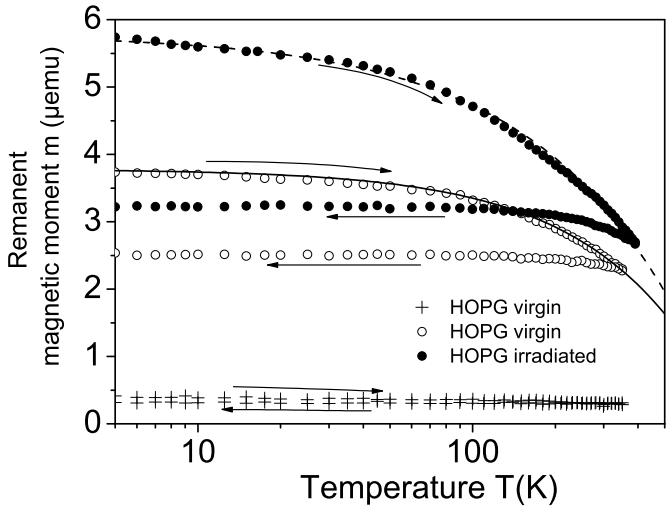}
\end{center}
\caption{Remanent moment vs. temperature of: $(\circ)$ virgin HOPG
sample of mass 14.2~mg. The continuous line is the function $m
[\mu$emu$] = 3.78 - 0.43 \times 10 ^{-2} T$ with $T$ in Kelvin.
$(+)$: virgin HOPG sample of mass 11.1~mg. $(\bullet)$: HOPG sample
(mass 14.2~mg) irradiated with low proton fluence ($\lesssim
0.1~$nC/$\mu$m$^2$) with a broad proton beam.  The dashed line
follows the function $m [\mu$emu$] = 5.78 - 2.65 \times 10^{-2}
T^{0.8}$.} \label{figrema}
\end{figure}

Figure~\ref{figrema} shows the remanent magnetic moment (zero
external field) of three HOPG samples (two virgin $(\circ,+)$ and one
irradiated $(\bullet)$) after magnetizing them with a field of 0.1~T
(1~T for the virgin sample $(+)$) at 300~K and cooling down in field
to 5~K. The finite remanence as well as the observed irreversibility
are clear signs for ferromagnetic behavior. The results of the two
virgin HOPG samples $(\circ,+)$ demonstrate that the magnetic
characteristics of nominally identical HOPG samples can be different
and stress the necessity of their characterization  prior to any
irradiation. According to  PIXE measurements there are no differences
in the magnetic impurity concentration between the three samples
shown in Fig.~\ref{figrema}. The observed differences should be
related to differences in the defect and/or hydrogen concentrations.

Recently published work showed that the magnetization of HOPG samples
after proton irradiation at low temperatures and at fields $B
\geqslant 0.1~$T decreases linearly with temperature \cite{barzola2}.
This behavior has been interpreted in terms of two-dimensional
Heisenberg model with a weak anisotropy. It is interesting to note
that the remanence of the virgin HOPG sample shown in
Fig.~\ref{figrema} shows also a linear temperature dependence whereas
the remanence of the irradiated HOPG sample a sublinear one. These
results indicate that the excitation of spin waves in magnetic
graphite does not follow the usual 3D Bloch $T^{3/2}$ model in
agreement with the results of Ref.~\cite{barzola2}. Note that both,
the irradiated as well as the virgin samples indicate a Curie
temperature above 400~K.

At the actual stage of knowledge and due to the narrow window of the
values of the irradiation par\-ameters necessary to trigger magnetic
order in graphite and the influence of external parameters (sample
dimensions, thermalization, etc.) one should not expect high
reproducibility between irradiation studies. The CMAM group (Madrid),
irradiated different HOPG samples with protons at 3~MeV energy and
observed a relatively large increase of the ferromagnetic moment at a
fluence $\sim 0.2~$nC/$\mu$m$^2$ \cite{ramos07}. Because the magnetic
moments of the selected virgin samples were not characterized before
irradiation, no quantitative estimate for the irradiation effect
could be done. Xia et al. \cite{xia} irradiated HOPG samples with
C$^+$-ions of 70~keV energy obtaining a similar behavior for the
ferromagnetic moment vs. fluence as previously observed
\cite{pabloprl03}. They obtained a linear temperature dependence for
the remanence of the irradiated samples and estimated a critical
temperature of 450~K \cite{xia} in agreement with the previously
reported behavior \cite{barzola2} and that shown in
Fig.~\ref{figrema}.

\section{XMCD results on magnetic spots produced on carbon films}

In Ref.~\cite{ohldagl} the spin dependence of the x-ray absorption
was studied in different proton-irradiated spots produced on two
different carbon films. The overall results in the two samples
indicate that the spin polarization of the pi-electrons leads to
magnetic order at the irradiated spots only. The surroundings of the
spots, a disordered carbon matrix, is para- or diamagnetic and do not
reveal any magnetic dichroism within experimental error at room
temperature. The XMCD measurements provide evidence for magnetic
order at the irradiated spots given by, for example,:\\
\noindent - the observation of magnetic domain patterns. These patterns depend on the
graphitized state of the film. \\
\noindent -   The appearance of magnetic anisotropy. The contrast
observed by XMCD on the partially graphitized
film indicate that the magnetization exhibits a preferred alignment  parallel to the film surface.\\
\noindent - The quantitative observation of a magnetic moment that
cannot be explained by paramagnetism or any of its related phenomena
at room temperature.

X-ray absorption at the Carbon K-edge only probes the electronic
structure of the 2p final state, because only direct optical
transition from 1s core level electrons into 2p final states
contribute to the absorption signal. The two main features in the
carbon absorption spectra  come from the two different types
(symmetries) of the s-p-orbitals that are possible and labeled
``$\pi$" and ``$\sigma$". Due to the negligible core-hole interaction
in carbon,
 intermediate excited states are not
observed in the K-edge soft x-ray absorption, thus the line shape is
only determined by the carbon $\pi$-states in the ground state
\cite{sto06}.

\section{Kelvin Probe Microscopy measurements}
As discussed above, proton irradiation in graphite samples can
produce a ferromagnetic phase in a specific part of the sample
interior only.  In particular for the case of ferromagnetic spots of
micrometer size \cite{barzola1,barzola2,ohldagl} the localization of
the magnetic part does not allow for an easy characterization of its
electrical properties. In order to characterize the electrical
properties of these spots Kelvin probe force microscopy measurements
on micrometer small magnetic spots have been performed \cite{sch08}.
The results reveal a charge transfer from the center of the spot to
an external ring, similar to that observed by XMCD \cite{ohldagl}.
Scanning transmission X-ray microscopy measurements on similar spots
reveal similar charge distribution. The potential variation of the
order of 50~mV and its distribution can be well understood in terms
of practically unscreened potentials \cite{sch08} indicating that the
magnetic material has insulating properties in agreement with the
expectations \cite{pis07,yaz08}.

\section{Magnetoresistance measurements}
A direct, alternative method to detect and study magnetic order is to
measure the magnetoresistance (MR). The MR develops a characteristic
butterfly loop when measured vs. magnetic field. Additionally, from a
ferromagnet we might expect that the MR depends on the orientation of
the magnetization with respect to the electric current direction,
i.e. the so called Anisotropic MR (AMR). We have prepared a HOPG
sample of dimensions $4.4 \times 1 \times 0.01~$mm$^3$ and irradiated
it with 12 spots, 0.8~mm diameter each, of 2~MeV protons of nominal
0.1~nC/$\mu$m$^2$ fluence.  Our four-point van der Pauw arrangement
and experimental conditions allow us low-noise transport measurements
with a relative resolution of $\sim 10~$ppm in the resistance $R$.

The non-irradiated HOPG sample showed basically no MR for the
parallel-to-the-planes field direction in agreement with earlier
results \cite{kempa03}. In that work it was demonstrated that the
always positive MR measured for this field direction is directly
proportional to the perpendicular field component only. Because our
HOPG samples are composed by crystallites with an average deviation
angle of $\sim 0.4^\circ$ respect to the nominal $c-$-axis of the
sample, a perfectly parallel field to the graphene planes cannot be
applied to the whole sample and this is the origin for the small and
positive MR.

\begin{figure}
\begin{center}
\includegraphics[width=0.5\textwidth]{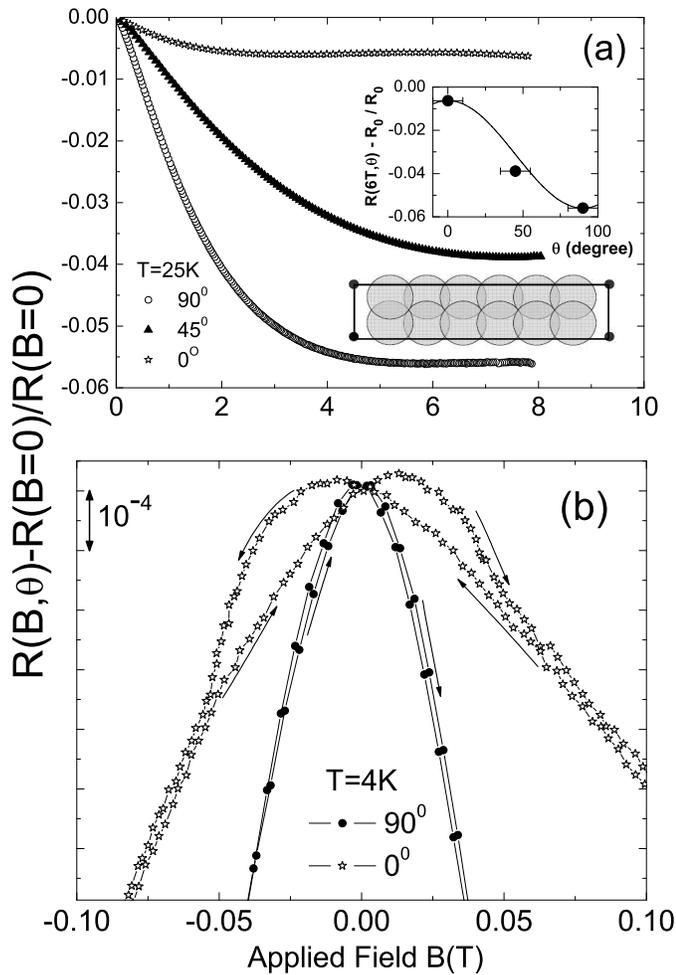}
\end{center}
\caption{(a) Magnetoresistance (MR) vs. magnetic field at 25~K at
three angles between field and current of a irradiated HOPG sample
with 12~spots (see sketch). The error in the angle $\theta$ was $\pm
10^\circ$. The inset shows the MR ($R_0 = R(0,\theta)$) vs. $\theta$
and the continuous line follows a $\cos^2(\theta)$ dependence. The
magnetic field is always parallel to the main area of the sample,
i.e. to the graphene layers. (b) Hysteresis loops in the MR for the
same sample. The measurements were done starting at a field of +8~T.
The asymmetry in the hysteresis at $\theta = 0^\circ$ might be to
different domain structures between the positive and negative
direction of the applied field and/or the contribution of a Hall-like
signal.}\label{MR}
\end{figure}

After irradiation the HOPG sample showed a negative MR and clear
hysteresis loops at low temperatures and fields confirming the
ferromagnetic state of HOPG after irradiation, see Fig.~\ref{MR}. The
irradiated sample had 100 times larger resistivity than in its virgin
state and showed a strong negative temperature coefficient indicating
a semiconducting-like behavior. Figure~\ref{MR}(a) shows the MR at
25~K at three angles $\theta$ between the field and the external
input current ($\theta = 0$ means field parallel to the current and
the sample main length). The observed behavior as well as the
absolute change of the MR is compatible with the AMR effect as, for
example, measured in ferromagnetic Co-nanowires \cite{dum02}. We note
that a positive and small $B^2$ contribution to the MR and observable
at fields above $\sim 5~$T has been subtracted from the data shown in
Fig.~\ref{MR}(a). In the upper inset of Fig.~\ref{MR}(a) the MR at
6~T field vs. angle $\theta$ follows the expected $\cos^2(\theta)$
dependence.

One may speculate that the AMR observed in the irradiated sample may
be due to the paramagnetic state and not to a ferromagnetic state of
the sample. Note that paramagnetic anisotropic MR has been observed
in, for example, SrRuO$_3$ at 170~K above the corresponding $T_C =
150~$K \cite{gen04}. However, several facts appear to rule out this
possibility. Measurements at different temperatures indicate that the
AMR effect decreases with temperature with a dependence incompatible
with a paramagnetic state (note that SQUID measurements show that the
paramagnetism produced by irradiation follows the Curie $1/T$ law
\cite{barzola1,barzola2,hoh08}). From the AMR data one can estimate
the change of the magnetization at saturation $M_s$ assuming that
$M_s \propto~(\textrm{MR})^{1/2}$ at high enough fields. The obtained
$M_s$ decreases linearly with $T$ in agreement with SQUID results in
other, irradiated and non-irradiated ferromagnetic, graphite samples
(see Sec.~\ref{squid}). These results indicate a critical temperature
$T_C = 190 \pm 10~$K for the sample shown in Fig.~\ref{MR}. Taking
the simple model from Ref.~\cite{pis07} and this $T_C$ we expect a
distance between vacancies between 2.5 and 3~nm, in fair agreement
with the estimated distance in the first $10~\mu$m depth for the
nominal fluence, taking into account the errors involved in the Monte
Carlo simulations and the lower proton energy used.

There is another experimental fact that speaks against an anisotropic
paramagnetic MR. Figure~\ref{MR}(b) shows the hysteresis measured in
the MR at low fields and at 4~K for two angles. This behavior
indicates ferromagnetic behavior and reflects the micromagnetic state
of the sample with the maxima expected at the coercive field $B_c
\sim \pm 0.013~$T, in agrement with SQUID results on other irradiated
samples.

The remarkable observation of the AMR in irradiated graphite
indicates a non-spherical symmetry of the charge distribution and a
non-negligible spin-orbit interaction. The anisotropy of the MR
should reflect the anisotropy in the electronic wavefunctions
\cite{ebe96} but several details have to be clarified in future
studies. To answer whether the interfaces between magnetic and
non-magnetic regions or only the domain behavior within the magnetic
regions play a role, further studies are necessary especially in
homogeneously irradiated and thinner HOPG samples.

\section{Conclusion}

The whole available data indicate that irradiated (as well as some
virgin) graphite can show magnetic order above room temperature. The
experimental evidence from SQUID, XMCD, transport and to some extent
also MFM measurements does not provide any indications that these
magnetic signals are related to magnetic impurities. The experimental
evidence does not support a (super)paramagnetic state as origin for
the observed magnetization effects. Defects appear to play a main
role although the question of the influence of hydrogen with or
without defects remains still open. Certainly, ion irradiation of
graphite is an interesting method to study systematically the
influence of defects and its magnetic properties. Systematic
transport, $\mu$SR, XMCD and NMR studies should be performed in the
future to understand the irradiation effects and finally the origin
for the observed magnetic order.

On the other hand, the results obtained using a MeV accelerator with
all its restrictions and the rather narrow window of parameters to
induce magnetic order with large enough amount of sample mass, stress
the necessity to search for more comfortable production methods. New
possibilities to produce magnetic graphite appeared recently. One is
through the irradiation of carbon ions at 70~keV energy. Although the
penetration depth of those ions is small ($\sim 250~$nm) the obtained
magnetic signals are high, reaching a magnetization of $\sim
10~$emu/g \cite{xia}. An alternative, peculiar method to produce
magnetic carbon is the one used in Ref.~\cite{par08} where magnetic
carbon powder was produced by a pulsed arc ignited between two carbon
electrodes submerged in ethanol, reaching a saturation magnetization
of $\sim 1~$emu/g. Instead of producing defects in highly oriented
graphite samples, other possibility to trigger magnetic order could
be to start from a disordered polymer with only C, O and H elements.
By annealing one removes O and H graphitizing partially the rest
carbon, which matrix remains highly disordered and may have
ferromagnetic regions.


\end{document}